# Mapping the German Diamond Open Access Journal Landscape

Authors
- Niels Taubert[1] (ORCID: 0000-0002-2357-2648)
- Linda Sterzik[1]
- Andre Bruns[1] (ORCID: 0000-0002-2976-0826)

[1] Institute for the Interdisciplinary Studies of Science (I²SoS), Bielefeld University, p.o. box 10 01 31, 33501 Bielefeld, Germany

Abstract

In the current scientific and political discourse surrounding the transformation of the scientific publication system, significant attention is focused on Diamond Open Access (OA). Diamond OA is characterized by no charges for readers or authors and relies on monetary allowances and voluntary work. This article explores the potential and challenges of Diamond OA journals, using Germany as a case study. Two key questions are addressed: first, the current role of such journals in the scientific publication system is determined through bibliometric analysis across various disciplines. Second, an investigation is conducted to assess the sustainability of Diamond OA journals and identify associated structural problems or potential breaking points. This investigation includes an in-depth expert interview study involving 20 editors of Diamond OA journals. The empirical results are presented using a landscape map that considers two dimensions: 'monetized and gift-based completion of tasks' and 'journal team size.' The bibliometric analysis reveals a substantial number of Diamond OA journals in the social sciences and humanities, but limited adoption in other fields. The model proves effective for small to mid-sized journals, but not for larger ones. Additionally, it was found that 23 Diamond OA journals have recently discontinued their operations. The expert interviews demonstrate the diversity within the landscape and the usefulness of the two dimensions in understanding key differences. Journals in two of the four quadrants of the map exemplify sustainable conditions, while the other two quadrants raise concerns about long-term stability. These concerns include limited funding leading to a lack of division of labor and an excessive burden on highly committed members. Gift-like contributions, while appealing, also present challenges as potential donors not only decide whether to contribute but also how to contribute, potentially creating friction between the gift and the journal's requirements. Furthermore, journals in the lower right quadrant often rely on third-party funding, necessitating a transformation once the funding expires. Common pathways for sustaining operations include lobbying for funding at the journal's home institution or increasing reliance on gift-based completion of tasks. These findings underscore the need for the development of more sustainable funding models to ensure the success of Diamond OA journals.

Keywords

Declarations

We declare no conflicts of interest.

Funding acknowledgement

This work was supported by the Federal Ministry of Education and Research (BMBF) and is a result of the project 'Community driven open access journals between scientific and economic requirements (CODRIA)' (Grant: AKZ 16TOA001).

In part, the empirical study was supported via the German Competence Network for Bibliometrics funded by the Federal Ministry of Education and Research (Grant: 16WIK2101A).

# 1. Introduction

In a recent announcement, the EU research council heralded a change in their objectives and priorities regarding Open Access (OA) that can be viewed as a paradigm shift: after decades of support for secondary publication (Green OA),[1] the development of infrastructures processing publication fees within research institutions[2] and the negotiation of large OA transformative agreements on the level of many member states[3], the tides now seem to shift towards OA that is free of charge for authors:

> "The EU is ready to agree that immediate open access to papers reporting publicly funded research should become the norm, without authors having to pay fees, and that the bloc should support non-profit scholarly publishing models. In a move that could send shockwaves through commercial scholarly publishing, the positions are due to be adopted by the Council of the EU member state governments later this month. […] To tackle inequalities in the ability of researchers to pay for publication, the text says that such fees should simply not be paid by authors and that non-commercial publishing models should be supported."[4]

The announcement is ambitious as tremendous effects are put in motion, promising a solution to both the problem of exclusion of scientists from accessing publications as well as the problem of ever raising costs for publishing research (Douglas, 1990; European Commission, 2006; Khoo, 2019). Journals that are addressed as a means for transforming the communication system of science are often called Diamond (Fuchs & Sandval, 2013) or Platinum journals (Haschak, 2007), and neither of them charge the authors nor do they request fees from the readers. Without doubt, such journals are likeable, as they seem to offer a non-commercial alternative to traditional publishing and seem more closely related to the scientific community as terms like scholarly-led (Edelmann & Schoßböck, 2020) suggest. However, before having high hopes regarding the impact of such journals on scholarly publishing and the for-profit publishing industry, a least two types of questions have to be answered. First, it is necessary to assess the current role of such journals in the communication system of science. How many of such journals exist and are these journals niche phenomena in scholarly publishing or do they publish larger parts of the research output of a country? Moreover, in which scientific fields could such journals be established? Second, it should be investigated whether the journals are sustainable or associated with structural problems or even breaking points. Given that they are defined in a negative way by the criterion of the absence of fees on the side of the authors and readers, this raises the question of how the journals manage to acquire the resources for their operation. The literature shows that Diamond OA journals receive their resources via different channels, including voluntary work of scientists, funds from institutions, third parties, (scholarly) associations and consortia as well as crowdfunding or a so-called freemium model where only part of the content is free of charge (Bachmann et al., 2022). Such a plurality already suggests that the journals are more diverse than terms like 'Diamond' or 'Platinum' evoke and that the question about the sustainability of the journals and possible built-in problems cannot be answered in a unified manner.

---

[1] Support instruments are the build-up and development of a repository infrastructure supported by the DRIVER and OpenAire projects (https://www.openaire.eu/history, accessed May 22, 2023), and the introduction of OA mandates in the 7th Framework Programme for Research, Horizon 2020 and Horizon Europe, in which immediate Green OA is accepted as one route to comply with the mandate. For the recent program Horizon Europe, see https://www.openaire.eu/how-to-comply-with-horizon-europe-mandate-for-publications (accessed May 22, 2023).

[2] As of May 22, 2023 the OpenAPC database lists 399 institutions worldwide that paid article processing charges for their authors (https://treemaps.openapc.net/apcdata/openapc/). Given that the data base is not exhaustive, it can be assumed that the number of research institutions that have access to publication funds is higher.

[3] For the three largest publishing houses, see: https://www.springernature.com/de/open-research/institutional-agreements, https://www.elsevier.com/open-access/agreements, https://www.wiley.com/en-us/network/publishing/research-publishing/editors/enabling-open-access-through-transformational-agreements (accessed May 22, 2023).

[4] https://www.researchprofessionalnews.com/rr-news-europe-infrastructure-2023-5-eu-ready-to-back-immediate-open-access-without-author-fees/ (accessed May 22, 2023).

The article takes these questions as a starting point and focuses on the German Diamond OA landscape. The case of Germany is of particular interest as the institutional research landscape is diverse (Powell & Dusdal, 2017; Hobert et al., 2021) and OA has been one of the priorities in research policy for a number of years. It can therefore be expected that the German Diamond OA landscape consists of a large variety of OA journals and allows us to gather insights into the situation of diverse journals.

To answer the questions raised above, the article will follow a two-folded strategy. First, the role of such journals in the communication system of science is determined by results of a quantitative bibliometric analysis that explores the landscape according to formal attributes. Second, a map of the field is developed in which the empirical evidence of a qualitative interview study with editors of German Diamond OA journals is organized. The map aims to shed some light on the diversity of the Diamond OA landscape and shows that the location of a journal on the map reflects positions of journals that are associated with leeway, challenges and problems the editors face.

## 2. Literature Review

The term 'Diamond OA' was invented by Christian Fuchs and Marisol Sandoval when disputing the adequacy of the term Gold OA as coined by Peter Suber (2012). According to Fuchs and Sandoval, the Gold OA category was considered as too broad as it does not distinguish between for-profit and non-profit OA journals. Therefore, they suggested a new OA category:

> "In the Diamond Open Access Model, not-for-profit, non-commercial organizations, associations or networks publish material that is made available online in digital format, is free of charge for readers and authors and does not allow commercial and for-profit re-use" (Fuchs & Sandoval, 2013, p. 438).

The fact that other authors picked up this suggestion (Costello, 2018; Leeds-Hurwitz, 2019; Rosnay, 2021; Cessna, 2023) indicates that the need for more differentiated categories was also shared by other scholars in the field. However, the use of three different criteria to distinguish Diamond OA from other types of journal-based – or in their words– corporate OA is a problem of the definition. The criteria point to the absence of financial interests on the side of the publishing entity, refer to the absence of payments for authors, and address license regulations that exclude any commercial re-use of the publications (Dellmann, et al.; 2022). The use of multiple criteria instead of a single one may lead to the need to invent further OA categories in cases in which not all criteria are met. For the purpose of this study, we will therefore use a pragmatic definition that is in line with Bosmann et al. (2021) and identifies with the term 'Diamond OA' journals that neither charge authors nor readers.

An overview of the literature on Diamond OA shows that studies on this topic are rare, but some aspects have already been investigated:

- Some studies ask for the *size of the Diamond OA landscape*. Suber's (2013) analysis is based on a study of the Directory of Open Access Journals (DOAJ).[5] It also includes information about possible fees and reports a fraction of 31% of the OA journals that apply article processing charges, while the larger majority abstains from charging authors with publication fees. Fuchs & Sandoval (2013) calculate smaller shares of the adaption of the APC model for the social sciences and the humanities that vary between 2.3% to 28.1%. Bosman et al. estimate the worldwide number of Diamond OA journals with a lower bound of ~17,000 and an upper bound of ~29,000 journals (Bosman et al., 2021, p. 27), while the number of articles published in them is estimated at 356,000, which is 8-9% of the worldwide scholarly publication output (ibid. p. 30). The magnitude is confirmed by a case study of Norway in which an increase of the Diamond OA output from 5 to 8% between 2017 and 2020 is reported (Frantsvåg, 2022). The study by Hahn et al. (2022) identifies 186 Diamond OA journals for Switzerland. Given that the methodology and the data sources that are used in the studies differ, it is not possible to directly compare the numbers that are reported here. Regarding regional distribution, a main focus for Diamond OA journals is Europe with 45% of the Diamond OA journals and Latin America with 25% (Bosman et al. 2021).
- A second aspect deals with the *size of Diamond OA journals,* and there is some evidence that they tend to be small to mid-size. On the worldwide level more than 50% of the journals publish less than 25 articles per year, while the share of journals with a yearly publication output >100 articles is only 3.5% (Bosman et al., p. 36). Large journals with more than 500 articles per year account for a share of not more than 0.2% of the worldwide Diamond OA landscape (ibid.). Regarding the average size, the study by Hahn et al. (2022) is in line with the findings that are reported for the global level.
- The *distribution of Diamond OA journals over disciplines or scientific fields* is another aspect for which empirical evidence is at hand. On the global level, social sciences and humanities make up around 60% of the Diamond OA journal landscape (ibid. p. 34). For Switzerland, Hahn et al. (2022, p. 12) report a share of more than 65% of journals in social sciences and Art and Humanities.
- In addition to such formal characteristics, various other aspects are studied. A survey with 1,609 representatives of Diamond OA journals shows that the Diamond OA landscape differs with respect to ownership, editorial work and quality insurance, disciplinary cultures, and organizational structure. A correspondence analysis underpins the diversity of Diamond OA journals as it results in the identification of the following five journal profiles that can frequently be found in the global Diamond OA landscape: 'small voluntary-run journals', 'learned society journals', 'institutional journals', 'publisher journals', and 'large professional journals' (Bosman et al., p.104-5). The impression of the heterogeneity of the field can also be derived from the study by Hahn et al. (2022). Aside from various other aspects, the study finds that Diamond OA journals vary with respect to the publishing languages, the publisher types, the licenses they apply and the type of peer review that is used.

---

[5] https://doaj.org/ (accessed May 22, 2023).

Besides the above-mentioned studies and opinion papers (e.g., Leeds-Hurwitz, 2019; Normad, 2018), a number of publications can be found that are written by editors of Diamond OA journals. Such publications do not report empirical findings but first-hand experiences about calls for the founding of new Diamond OA journals (e.g. Lefebvre 2022), reflections on the founding of the journal (e.g., Bachmann et al. 2022) or the organization of the processes and the costs for the journal (e.g., Rehmann 2003). Such publications provide valuable insights and an important background for the understanding of such journals. Against the background of the literature, the aim of this article can be characterized as original for two reasons: regarding the role of such journals in the communication system of science, this has been investigated on the global level and for Switzerland but not for Diamond OA journals that are located in Germany. What is much more important, however, is that it takes a resource-oriented perspective and asks for the journals` positions in the field and possible problems that are associated with them.

## 3. Methods

The analysis is based on a mixed methods approach that combines three components: a bibliometric analysis of the Diamond OA landscape, a survey with a selected sample of editors of 20 Diamond OA journals and semi-structured expert interviews with the same group of editors.

### 3.1 Bibliometric analysis

For the bibliometric mapping of the German Diamond OA landscape a quality controlled, exhaustive list of Diamond OA journals was created from March 2022 to July 2022. For the identification of German journals, a number of criteria were considered, including the location of the publishing organization, the funding of the journal by a German organization or the location of the editorial office. However, these criteria may lead to an ambiguous attribution of journals as publishing organizations and editorial offices may be located in more than one country and a journal may receive funds from more than one funding organization in different countries. Therefore, it was decided to use the journals' hosting location as mentioned on their website as a demarcation criterion.[6] Three approaches were used as sources for the identification of possible Diamond OA journals.

- Since many Diamond OA journals are hosted by Open Journal Systems (OJS) platforms, a list of OJS-hosting services in Germany[7] was taken as a starting point for manual data collection. Websites of all hosting services were visited, and basic information like journal names and ISSN were collected.
- The second approach aimed at identifying the sub-group of Diamond OA journals in world-wide evidence sources for Gold OA journals. An initial list was compiled[8] which included all serials with ISSN from the Directory of Open Access Journals (DOAJ)[9], PubMed Central (PMC)[10], and the Directory of Open Access Scholarly Resources (ROAD)[11]. In a second step, data from DOAJ, PMC and ROAD was searched for country information to identify journals hosted in Germany as well as information about whether or not any fees are charged from authors to identify journals that are 'Diamond'. In the case of ISSN covered by DOAJ, the list was restricted to ISSN with the country information 'Germany' and the entry 'No' in the field 'Publication fees'. In cases where the string in the field 'publisher' and/or 'institution' comprised a keyword that indicated a German city or institution (e.g., 'Universität', 'Hochschule', 'Institut', as well as the names of German university cities), the ISSN was also included. For ISSN covered by PMC a similar procedure was applied. For ISSN included in the ROAD dataset, all ISSNs were obtained where the website contained the top-level domain '.de' or the publisher string contained one of the mentioned keywords.

---

[6] For a more inclusive demarcation strategy, see Hahn et al. 2022.
[7] https://ojs-de.net/netzwerk/ojs-standorte-im-deutschsprachigen-raum (accessed May 22, 2023).
[8] The resulting ISSN Gold-OA-List is published (https://pub.uni-bielefeld.de/record/2961544, accessed May 22, 2023).

- After the announcement of the first version of the list (DOAG 1.0), the community of German librarians provided valuable feedback and pointed to possible additional Diamond OA journals. These journals were added to the collection of possible candidates.

The journals identified by the three approaches were merged in a single list, followed by a verification process. For this purpose, the OA status of the journals was verified by matching them against the ISSN-Gold-OA list 5.0 (Bruns et al. 2022), which is a curated evidence source for Gold OA journals. In addition, the journals for which payment information were found in OpenAPC (Pieper & Bronschinski 2018)[12] for 2020 and 2021 were removed from the list as they charge a certain kind of fee. To assure the correctness of the Diamond status, a manual checking procedure was applied: the journal websites were visited and searched for information as to whether or not the journals charge any publication fees. In cases of missing or unclear information, the editor of the journal was contacted and asked for clarification. During the manual checking procedure, it turned out that the list covers not only journals but also a number of other types of media. To identify them, such media were flagged and the final list was published (Bruns et al. 2022b). For the purpose of this study, further information were collected from the journals` websites, including the number of publications published in 2021, the number of articles in 2021, the type of peer review that is applied, and the publisher entity. Based on the journals` descriptions (typically in the 'aims and scope'-section), all journals were manually assigned to one of the six large scientific fields as defined by OECD category scheme.[13]

### 3.2 Survey with editors of 20 Diamond OA journals

To study the operation of Diamond OA journals in-depth, a survey and interviews with editors of 20 Diamond OA journals were performed. The interview sample aims to represent a maximum variation (Suri 2011, p. 67) of the Diamond OA landscape in terms of subject field in which the journal publishes, size as measured by the numbers of publications, and models of acquiring resources.

---

[9] https://doaj.org Data download: 29.09.2021). Currently there are 17,570 journals listed in the DOAJ, of which 12,074 (~ 68.72%) do not charge any article processing charges (APC).
[10] https://www.ncbi.nlm.nih.gov/pmc/journals/ (Data download: 29.09.2021).
[11] https://www.issn.org/the-issn-international-is-pleased-to-introduce-road/ (ROAD data can be accessed after registration. Data download: 29.09.2021).
[12] OpenAPC is a large collection of information about actual payments for article processing charges. At the time of writing, it includes payment information from 21 countries (https://openapc.net/, accessed May 22, 2023).
[13] http://help.prod-incites.com/inCites2Live/filterValuesGroup/researchAreaSchema/oecdCategoryScheme.html (accessed May 22, 2023).

Given that we were interested in both a number of factual information about the journal as well as narratives and interpretations from the editors, such as the history of the journal, the development of its reputation and its actual position in the field, it was decided to collect the information separately. For the factual information and for the preparation of the interviews, an online survey with closed questions was created and supported by the LimeSurvey.[14] The questionnaire included questions about the journal (e.g., year of founding, the year of application of the Diamond OA model, journal ownership), questions about financial support (funding), division of labor in the editorial offices, distinction between paid and unpaid tasks, possible cooperating partners and the technical infrastructure (e.g., editorial management system, journal platforms) used by the journal. The survey was filled out by the interviewees usually a couple of days up to few weeks before the interview was conducted. 19 of the 20 journal editors completed the questionnaire, a task that took about ten minutes.

### 3.3 Interviews with editors

Guideline expert interviews that were conducted with the editors of the sample are the third component of the methodological design. The aim of the interviews was to gather narratives about the founding of the journal, its relations to the scientific community, the funding history, the everyday operations of the editorial offices, the distinction between paid and unpaid work, the leeway of and restrictions the editorial team faces in the development of the journal, as well as collaboration with third parties and experiences with the infrastructure that is used. Given that the interviews addressed internal matters of the journals, and given that past experiences indicate that such information are regarded as sensitive by the interviewees (Morrison, 2016), anonymity was guaranteed. The interviews were conducted via the online conference platform Zoom between March and July 2022 and were recorded. Regarding length, the interviews varied between 28 minutes and 173 minutes. All interviews were transcribed by a transcription service to guarantee the quality of the transcriptions. For all interviews a qualitative content analysis was performed following the principles in Mayring (2015, 2019). To this end, a code tree was developed with 181 different codes, and all relevant interview passages were assigned to a code by hand. When the analysis was completed, more than 2,600 text passages were coded. The coding and further analysis was supported by MaxQDA 2022.

## 4.The German Diamond OA landscape in numbers

In this section, the German Diamond OA landscape is described by presenting the results of a quantitative analysis of the journals covered by the Diamond OA List Germany (DOAG). DOAG contains 659 distinct ISSN and 458 distinct journals. Each ISSN and journal can be covered in one or more of the data-sources. Numbers were calculated after manual data cleaning. 418 of the ISSN (~63.4%) and 323 of the journals (~70.5%) are indexed in ROAD. About 36.7% of all ISSNs and 37.3% of all journals are covered in DOAJ. About 40.6% of the journals are included in the manually collected data set from OJS hosting services. For our analysis, six publication media were excluded from the list as it turned out that they were hosted in another country.

As a first result, we would like to highlight that the data collection does not only yield a list of scientific journals but a more diverse set that also includes other media than journals. Figure 1 shows the description of the media types with different types of series (such as technical reports, working paper series), internal publications that are open to a restricted group of authors (in most cases from a particular institution), student journals that publish term papers and other works by MA and BA students[15], blogs that address a wider public, conference proceedings and practitioners` journals that provide information for a professional field of practice (e.g., lawyers, engineers, or teachers). The further analysis is limited to the 298 Diamond OA scientific journals.

---

[14] https://www.limesurvey.org/ (accessed May 22, 2023).

*Figure 1*: Journals and other publication media in DOAG

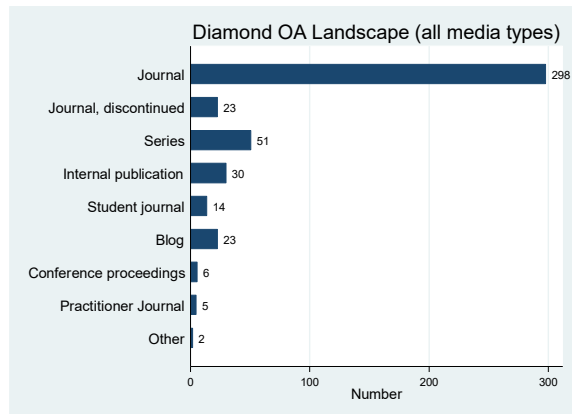

It is worth noting that 23 journals did not publish any publication in 2020 and 2021 and were therefore regarded as discontinued. This number indicates that there is some dynamic in the field and that there are not only new Diamond OA journals entering the field (Hahn et al., 2022) but also a remarkable number that stopped operating.

Second, and when turning to the distribution of German Diamond OA journals over the six major scientific fields of the OECD category scheme,[16] two main fields can be identified, to which the majority belongs. Figure 2 shows that more than 70% of the Diamond OA journals are either assigned to the social sciences or the humanities. This is in line with previous findings (Hahn et al. 2022, p.12; Bosman et al. 2021, p. 34).

---

[15] For the relevance of student journals for students' engagement with Academy, see Uigin et al., 2015.
[16] https://www.oecd.org/science/inno/38235147.pdf (accessed May 22, 2023).

*Figure 2*: Diamond OA journals by OECD major fields

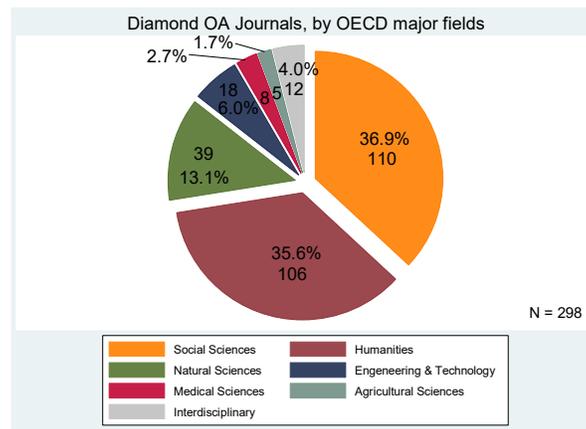

Natural sciences (including a number of journals in mathematics) contribute 13% of Diamond OA journals, while the numbers in Engineering & Technology, Medical Sciences, and Agricultural Sciences are small. Regarding the publishing entity that is mentioned in the imprint of the journal, a focus can again be found. Figure 3 indicates that a majority of nearly 58% are published by a research institution, which includes universities but also the large variety of non-university research institutions in Germany. Commercial publishing houses are second in terms of numbers of Diamond OA journals published, 17% (or 51 journals in absolute numbers). This result may be a bit surprising, especially as Diamond OA is often associated with not-for-profit (or non-commercial) publishing.[17] Diamond OA journals published by commercial publishing houses often apply the subscribe-to-open model in which libraries keep up their subscriptions for well-established journals and do not pay for reader access to the journal but for OA to all content published by the journal. In situations in which the number of subscriptions falls below a certain threshold of subscriptions, the journal will return to a pay-for-reader access subscription model (Crow et al. 2020). Besides the two categories, learned societies act as imprints in more of 10% of the journals and 'other public institutions', 'individuals', and other publishing entities add smaller fractions between 4-6% of the journals.

*Figure 3*: Diamond OA journals by Publishing Entity

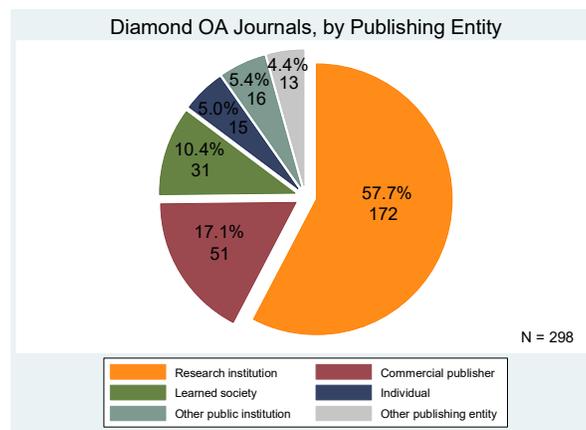

---

[17] Note that some of the publications of learned societies were published in cooperation with for-profit publishers and such journals were assigned to the category 'Commercial Publisher' due to the imprint on the website.

To determine the size of German Diamond OA journals in terms of publication output, we visited the journals' websites and collected the number of publications and articles in 2021 manually. In eleven cases, the website of the journal was down and the number of articles could not be investigated. For a contextualization, the results of the German Diamond OA landscape were compared with the group of journals that are indexed by the Web of Science (WoS) in-house database maintained by the German Competence Center for Bibliometrics (WoS-KB) in its 1/2023 version.[18] The main advantage of using this data source is that the raw data are available and allow a straightforward calculation of the publication output of all journals included. Even though the two sets of journals show many differences, the comparison with the WoS journals is instructive as the database covers the most relevant journals.

Regarding the group of Diamond OA journals, the descriptive statistics show some peculiarity: on the lower end, the minimum value of 0 suggests that some of the Diamond OA journals publish irregularly, while on the other end of the spectrum, the two journals with 1,149 and 2,834 are strong outliers. A further investigation reveals the causes: both journals - the *Journal of High Energy Physics* and the *European Physical Journal C: Particles and Fields* – are part of the SCOAP3 consortium in high energy physics (HEP), in which funding bodies and libraries worldwide pay centrally for the service of the publishers by re-directing funds that were formerly used for the payment of subscriptions (Mele et al., 2009). By accident, the two journals are part of the Diamond OA list as they are published by the imprint Springer Heidelberg, which is located in Germany. In total, the 3,983 articles that are published in the two SCOAP3 journals account for 33.0% of the 9,044 articles and 44.0% of the 12,086 publications (all publication types, including articles) and lead to distortions regarding descriptive statistics. Therefore, we will give both numbers including and excluding the two HEP journals.

The average number of publications published in German Diamond OA journals in 2021 is 42.11 (with SCOAP3 journals) or 28.43, respectively (without SCOAP3), as Table 1 shows. A comparison with the journals covered in WoS puts the number into perspective and makes clear that the average is 4.4 to 6.5 times larger for this group. The median of 16 publications and 11 articles points to the fact that half of the Diamond OA journal group have a small publication output, while in the WoS journal group the median of 72 publications or 59 articles indicates that there are also mid-size journals below the median. A large standard deviation for all publications and articles in all journal groups indicates that the size of the journals varies on a large scale. The probably most impressive difference refers to the journals with the largest article output. In the Diamond OA journal group (without SCOAP3 journals), the maximum article output is 288, while in the group of the Web of Science journals, the largest journal published no less than 23,307 articles in 2021.

---

[18] https://bibliometrie.info/ (accessed May 8, 2023).

*Table 1*: Publication and Article output of Diamond OA journals vs. journals in Web of Science (2021)

|  | Diamond OA journals | Diamond OA without SCOAP3 | Web of Science |
|---|---|---|---|
| *All Publications* | | | |
| Mean | 42.11 | 28.43 | 186.50 |
| Median | 16 | 16 | 72 |
| Std. deviation | 186.76 | 56.35 | 527,81 |
| Min. | 0 | 0 | 1 |
| Max. | 2,834 | 738 | 24,593 |
| Observations | 287 | 285 | 16.220 |
| *Articles* | | | |
| Mean | 31.51 | 17.76 | 153.62 |
| Median | 11 | 11 | 59 |
| Std. deviation | 180.75 | 25.69 | 473.39 |
| Min. | 0 | 0 | 0 |
| Max. | 2,834 | 288 | 23,307 |
| Observations | 287 | 285 | 13,453 |

Figures 4 and 5 provide a more detailed insight into the distribution of the article output of German Diamond OA journals. If we again exclude the two large HEP journals, 209 journals or 73.3% of the German Diamond OA landscape publish 20 or less publications. They account for only 33.6% of the articles. If we focus on the journals below the median of 11 publications, they published only 712 articles or 14.1% of the total article output of the German Diamond OA landscape in 2021. A look at the other side of the spectrum reveals that 40.5% of the article output in 2021 appeared in the largest 10% of the journals. Moreover, the 5 largest Diamond OA journals published 767 articles in 2021, which is more than the publication output of the journals that are below the median.

When compared with the group of journals covered by WoS, it becomes clear that such a skewed distribution of the publication output of journals also exists in this database (see Figure 6). Note that journals with a publication output of more than 1,000 articles were excluded for reason of presentation. However, a comparison of Figure 5 with Figure 7 shows that the skewness happens on a different level. The WoS also includes small journals with less than 20 articles but includes a massive share of 32.84% of journals with a publication output of more than 100 articles in 2021.

*Figure 4 & 5*: Distribution of article output and number of Diamond OA journals, grouped by article output

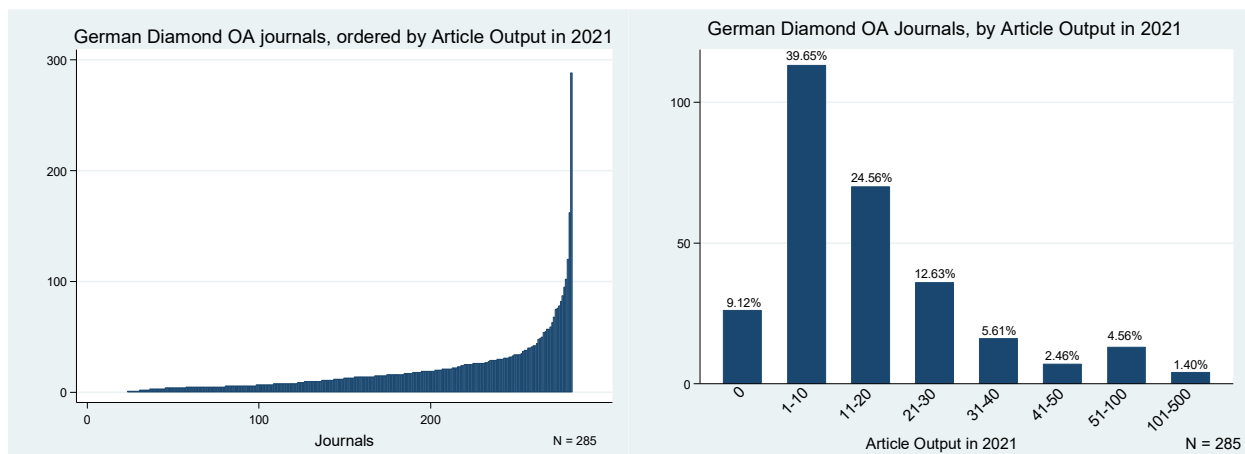

*Figure 6 & 7*: Distribution of article output and number of WoS journals, grouped by article output

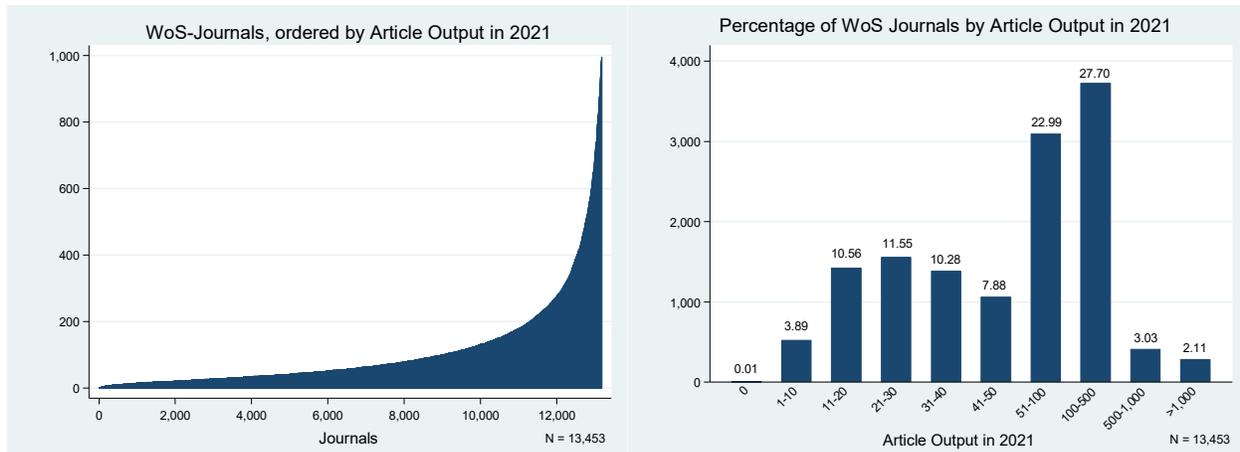

We will conclude with the quantitative analysis of the group of German Diamond OA journals by asking whether the size of the publication output is associated with other properties. Table 2 shows the size of the article output by OECD major fields. Three results are worth highlighting: first, the Diamond OA journals in the humanities have a smaller average mean article output than journals in the natural sciences and engineering and technology. The numbers for medical sciences and agricultural sciences should not be over-interpreted because of a small number of journals in these fields. Second, larger journals with an article output of more than 50 articles can be found in the natural sciences, engineering & technology, social sciences and agricultural sciences but not in the humanities. However, all scientific fields have Diamond OA journals with a small publication output. Third, the Diamond OA journals of the social sciences tend to hold a position in the middle, both with respect to the article output and the largest journals in the field.

*Table 2:* Number of articles (2021) in German Diamond OA Journals, *by OECD Major field\**

| OECD Major Field | Number Journals | Mean | Std. deviation | Min. | Max. |
|---|---|---|---|---|---|
| Social Sciences | 105 | 17.14 | 15.71 | 0 | 95 |
| Humanities | 103 | 9.53 | 9.58 | 0 | 44 |
| Natural Sciences | 35 | 38.4 | 50.81 | 4 | 288 |
| Engineering & Technology | 18 | 31.78 | 42.04 | 0 | 162 |
| Medical Sciences | 8 | 11 | 8.07 | 0 | 25 |
| Agricultural Science | 5 | 24.2 | 19.83 | 3 | 54 |
| Interdisciplinary | 11 | 14 | 25.33 | 0 | 87 |
| All | | | | | |

\* SCOAP3 HEP journals were excluded to avoid distortion

In Table 3, descriptive statistics are given for the journals' article output and are differentiated by the type of publisher. The results show that for all publisher types, journals that publish irregularly exist, as the minimum number of 0 s indicates. Mid-size journals with an article output or more than 50 in 2021 can be found for all publisher types with the exception of Diamond OA journals that are published by individuals. Such journals are always small with respect to the article output, which is also reflected by a small standard deviation. Commercial publishers tend to publish journals with larger article outputs, as indicated by the mean value.

*Table 3:* Number of articles (2021) in German Diamond OA Journals, by publisher type*

| Publisher Type | Number Journals | Mean | Std. deviation | Min. | Max. |
| --- | --- | --- | --- | --- | --- |
| Research Institution | 165 | 14.56 | 17.11 | 0 | 102 |
| Commercial Publisher | 49 | 28.65 | 42.15 | 0 | 288 |
| Learned Society | 31 | 21.81 | 28.62 | 0 | 162 |
| Individual | 14 | 5.36 | 3.46 | 0 | 11 |
| Other public institution | 13 | 16.08 | 17.28 | 0 | 54 |
| Other publishing entity | 13 | 22.62 | 35.78 | 0 | 120 |
| All | 285 | 17.76 | 25.69 | 0 | 288 |

* SCOAP3 HEP journals were excluded to avoid distortion

## 5. A two-dimensional model of the Diamond OA journal landscape

The bibliometric analysis already provides some evidence that also German Diamond OA journals are diverse. In the next step, we will take a resource-oriented perspective and map the Diamond OA journal landscape along two dimensions that turned out to be most impactful for Diamond OA journals during the analysis of expert interviews. The map aims to order the diversity of the field and shall help to understand the resource-related situation of Diamond OA journals. The first dimension describes to what extent tasks that are necessary to run a journal are monetized and to what extent they are based on gifts. The second dimension is the size of the journal team. Before turning to the empirical results of an in-depth analysis of a sub-sample of 20 Diamond OA journals, we elaborate on the two dimensions in more detail.

For the development of the first dimension, it is helpful to draw on the literature on the digital commons (Hess & Ostrom, 2007) and regard Diamond OA journals as a two-sided case. On the one side, content that is published in Diamond OA journals constitutes, like in all types of OA, a common good on the reader-side, as it provides free access to research published in the journal. On the other side, it also constitutes a common good on the author-side, as it offers the opportunity to publish research free of charge, provided that the manuscript fits into the scope and passes the selection procedure (e.g., peer review). With respect to the production, two types of resources are discussed in the literature: some observers highlight the relevance of a monetary income that comes from different sources including third party funds and scholarly associations (Gajovic, 2020) or outside donors more generally (Normand, 2018). Other scholars point to hard-working volunteers that are necessary to run a Diamond OA journal (Bamberg, 2012; Hoorn, 2014; Morrison, 2016; Hahn et al., 2022). Bosman et al. (2021, p. 8) report a share of "60% of OA Diamond journals that depend on volunteers" and identify at the same time a "wide range of funding mechanisms" that the journals apply. These findings suggest that monetary resources and voluntary work are not mutually exclusive alternatives but are often combined for resourcing Diamond OA journals. For the conceptualization of the first dimension, we will therefore not focus on journals but on entities of finer granularity. These are tasks that have to be completed for the journals' operation and we will distinguish between a *monetized and gift-based completion of such tasks.*

- *'Monetization'* indicates that work is done based on a payment. From the perspective of the individual, money acts as an incentive for action, while from the perspective of the journal, another aspect is more important: paid work is usually based on a contract (Bloch & Perry, 1989, p. 5) which expresses a mutual agreement that a defined task is attributed to an employee who is responsible for its completion. Therefore, the contract makes sure that the completion of a task is reliable and predictable for the journal.
- The *gift-based completion of tasks* follows a different logic. According to the seminal work of Marcel Mauss (1954) on non-monetized economies, gifts always bear something of the personality of the donor and are given and repaid under obligation. However, gifts to the digital commons differ in many respects from the exchange systems that Mauss studied. The most fundamental difference is that there is no exchange between certain identifiable groups or individuals (Taubert 2006). Therefore, gift-giving practices in the digital commons are not based on reciprocity. "The contributions made for the digital commons surely are gifts, but these are gifts to humanity, not to specific and selected people. They are gifts without an obligation to return" (Willer 2013). If no obligation exists, a possible donor is free to decide *whether or not* to contribute and, in many cases, also *how* to contribute.[19]

In the context of our study, the dimension 'monetized-/gift-based completion of tasks' was operationalized in the online survey with the 20 journal editors. A number of service-oriented tasks were defined, including the handling of manuscripts, organization of the correspondence, preparation of decision-making, proofreading, copy-editing, preparation of the layout of accepted journals, creation of metadata, notification of metadata, hosting of the editorial system, and the maintenance of the journal platform. In the survey, the editors were asked which of the tasks occur for their journal and whether or not the completion of the tasks is paid for. The information collected in the survey was validated against the narratives about the journal operation, which were provided by the expert interviews.

*Size of the journal team*: The second dimension is easier to develop. The primary focus is again on service-oriented tasks and the dimension describes how many people are involved in their completion. In combination with the dimension 'monetized/gift-based completion of tasks' the size of the journal team may imply different things at the two poles of the first dimension. At the pole of a low-monetization, small journal teams will result in the necessity that individuals contribute with large gifts to serve the journals' demands, while large journal teams may bear the opportunity that all members are contributing with small gifts, provided that the workload is equally distributed. At the pole of high monetarization, large teams may bear the opportunity for a larger degree of task specialization, while in small teams, the division of labor and the specialization have to be low. For the operationalization of the dimension 'size of the journal team', information that were collected from the journals' websites were validated against the evidence from the expert interviews. Figure 6 shows the map of the Diamond OA landscape for the 20 journals that were studied in-depth.

---

[19] In the empirical results an exception will be presented.

*Figure 8*: Map of the Diamond OA Journal Landscape

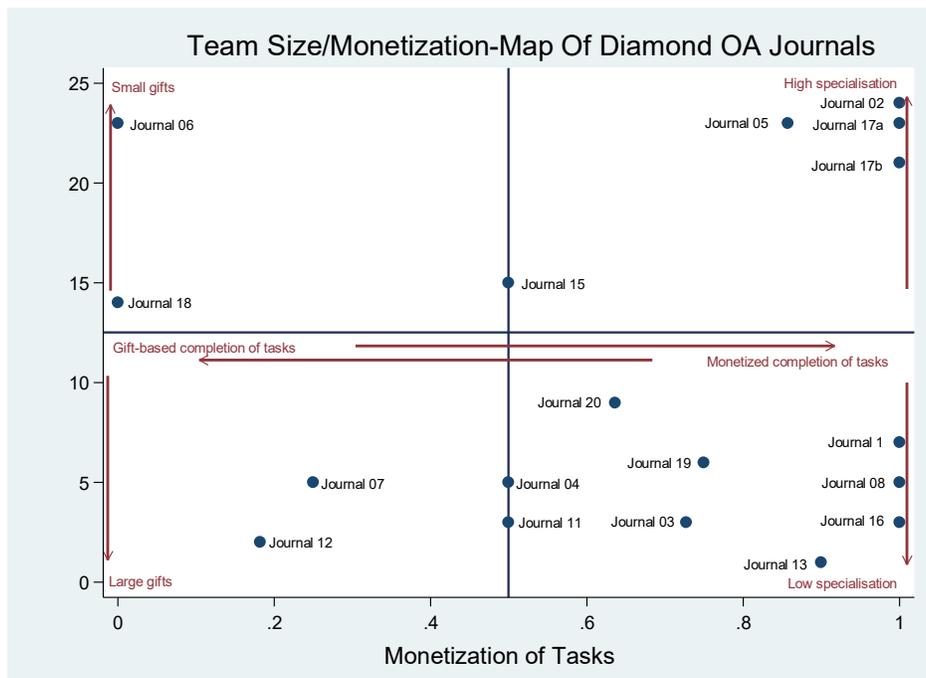

# 6. Positioning and its consequences – empirical results from an in-depth study

So far, the considerations about the map of Diamond OA journals are abstract in a sense that it is more based on logic reasoning and 'hidden' empirical evidence. In this section, we will fill this gap and provide some findings about the consequences that a position has for a journal. Our strategy is not to present one case after another but to choose a more problem-oriented presentation that refers to particular positions in the field.

## 6.1 Upper left quadrant: The miracle of the crowd

One of the probably most fascinating models of running a Diamond OA journal is to complete all tasks with a large journal team but without any payments involved. On the map such journals are located in the upper left quadrant. An example is J-06, a journal that covers a subject field in pure mathematics with an impressive three-digit yearly article output. The digital born journal was founded in the mid-1990s and provides free reading and publishing without charging authors and readers. Regarding its reputation, interviewee I-06 describes it as not being one of the two flagship journals in the field but considers it in third position in the reputation pyramid (I-06, pos. 64-78). The reason why such an innovative model occurred in mathematics is explained by the interviewee with reference to the foresightedness of the founding editor together with a specific cultural attitude of mathematicians:

"In this subject field there was enough space for a new journal. And there was Mr. W. [name of the founding editor] who sadly died a few years ago, who was very innovative. And he was at the right place to try out something new. Well, I mean in 1994 the Internet was in its infancy. And he has somehow seen the potential. And the other reason why it has been founded is […] that research mathematicians are a little bit anarchistic, they're not like the classic natural scientists, they're more of a bit of an artist. And that offers the background that such projects are accepted." (I-06, pos. 20-36)

In the quotation, the mindset of mathematicians regarding innovative journals is characterized with a triad of references: anarchism refers to challenging inherited (power) structures, the negative references to natural scientists are likely to point to a lesser orientation of mathematicians towards well established and prestigious journals of large publishers, and the figure of the artist appeals to creativity and a positive attitude towards novelty. The principles of the journal model were invented at the time of the journals` foundation and are still in place today.[20]

"Back then, the editors developed a model that I really like and that still exists. Well, that's an indication that it sustained for fifteen years or more. If you have a journal with a large turnover, a classical model with two, three or four editors-in-chief is not sustainable. […] If you have a large journal, the editors-in-chief are usually paid for. […] and this is missing for our journal. Well, and then they noticed "Okay, we'll have ten or fifteen editors-in-chief". […] So, it works that we have two managing editors and I am one of them. We receive the papers and distribute it to the editors-in-chief. 90% of my work is that I receive a paper, have a look at it and say, "Yes, that looks like as it could be well-suited to by managed by LS" [Name of a colleague]. And via our system LS receives a short email and from that point LS is de facto editor-in-chief. She will do everything until the paper is accepted or rather it is suggested for acceptance to the rest of the editorial board. And this model distributes the workload. Back then, it was something very, very innovative, but like I said it works." (I-06 pos. 144-172)

It would be a misconception to understand the organizational model of the journal as simply an involvement of a large number of mathematicians. Moreover, it is characterized by a specific division of labor that consists of two types of roles. The first one is the role of the two managing editors that act as a hub. They receive the submissions, overlook the topics and hand them to one of the editors-in-chief based on his or her competencies. The second one is the role of the editors-in-chief who differ with respect to their expertise but not regarding the tasks they perform. They are responsible for the management of the whole peer review process, including the selection of referees, the correspondence with referees and authors, the reading of the referee reports, the preparation of a decision, and the presentation of it to their colleagues. In other words, the distribution of work is not achieved by decomposing the editorial process into a number or separate tasks and by attributing them to particular team members, but via a distribution of incoming manuscripts so that each editor-in-chief is only responsible for a tiny fraction of them. In the discussion about peer production, it is highlighted that such a fine granularity of tasks is an important precondition for the division of labor (Benkler and Nissenbaum 2006). However, the miracle-of-the-crowd-model is not only a specific way of organizing the editorial work but is also accompanied by technical preconditions and a specific orientation of the journal editors. Both technical and institutional aspects lead to a minimization of the workload for all parties involved. The technical preconditions refer to a unique development in mathematics which is the software package LaTeX:

"We have our own time typesetting software, that is TeX or LaTeX now, it's been around since the 1980s, late 1970s. And that means that right from the start a very, very large task did not exist […] There were many journals sort of expected that mathematicians would do their own typesetting, so to speak. When a paper was accepted somewhere, then the publishers were still working on it back then, but the mathematicians were already doing the basic typesetting. And the H, W and Q [names of the founding editors], who then did it together, said: "Yeah well, if you do it anyway, there's this new thing now, the

---

[20] Evidence for an affinity of mathematics and the diamond OA model can be found in Teschke (2018) who reports a large number of such journals for this discipline.

Internet, then we can [...] design them ourselves. Yes. And with that, a major cost factor is simply gone and now let's try it like this for free, so to speak." (I-06, pos. 36-57)

As we will see, typesetting is a task a number of other Diamond OA journals struggle with. In the case of J-06, this is done by the provision of a LaTeX template and by assigning the task to the authors. Such organizational principle reminds us of the distribution of work under the editors-in-chief just discussed, as the same task is performed by a large number of people, instead of by a single specialized position that may have to be paid. The second feature that reduces the work is a pragmatic orientation. An example can be found in the way in which the decisions about how to proceed with a manuscript after review are organized. Such decisions can be regarded as the most significant and most symbolically valued in the every-day-operation of the journals. In the case of J-06, decision-making is simplified at maximum, as interviewee I-06 depict:

> "If one of the editors-in-chief say: 'we want to reject it,' it will be rejected. Such a doodle. If he or she says 'it should be accepted,' then the whole board has to vote, or respectively it is handled that one comes up and says: 'Hey I would like to accept it for this and that reasons.' And then we wait to see if anyone criticizes this decision. 90% of the time it doesn't happen and then eventually it's accepted." (I-06, pos. 201-211)

The interview passage makes clear that the pragmatic way of decision-making puts the respective editor into a powerful position. She can solely decide to reject a paper, and when it comes to acceptance, her suggestion is influential as the other editors are put in the position to object in the case that they disagree. In other words, the mathematician who did the editorial work for a particular manuscript has the most influence on the decision. A final aspect of the organization of work is a fair distribution of work.

> "I need enough people, who will really do the work without being paid. And if new people join us, we will tell them: 'Look out there is work that has to be done.' But this is our model and we reduce the workload to a minimum [...] These are details, but we are really attentive, we have our own brand sheet and we track that the workload is equally distributed." (I-06, pos. 179-205)

Fairness of the distribution of work implies that all editors-in-chief receive roughly the same number of manuscripts they are responsible for and contribute to the journal with gifts of roughly the same size. However, the quotation does not only address the fair-distribution-of work-norm but also points to another aspect that mathematicians have to agree with if they want to contribute to the journal. They have to be ready to accept the distribution of work, which means to be responsible for the management of the complete process and not only single tasks. In other words, they are free to choose whether or not to contribute to the journal, but if they decide to contribute, they are supposed to contribute with a specific type of gift.

One aspect that is not explicitly mentioned in the interviews but results as a logical consequence of the model is that reduction of the work for decision-making can only be achieved if there is mutual trust under the editors that the others make reasonable decisions they in principle agree with. Such behavior goes along with a surrender of control by one of the editors.

## 6.2 Upper right quadrant: Stable, well-funded journals

It may be coincidence or not, but the journals in the upper right quadrant are all journals in the natural sciences, including one journal from mathematics. They are characterized by a large degree of monetarization and relatively large journal teams and correspond to the 'large professional journal' type as suggested by Bosman et al. (2021, p. 1.04). The reasons for the focus in the natural sciences might have to do with the availability of resources in the disciplines. Two of the journals are funded by a foundation, one is generously supported by an institute of the Wissensgemeinschaft Leibniz (WGL), and one receives subsidies from a research institute. However, the journals of the quadrant do not simply have more monetary resources than other journals in our sample. Moreover, their funding is also not limited in time. One of the interviewees explains the relevance of such non-terminated funding as follows:

> "Basically, I don't think it's a favorable development or a favorable model that Diamond Open Access Journals receive funding, wherever the funding comes from. Something like this is always temporary. And in science in general, funding is almost always limited in time. And then the funding runs out and the whole thing collapses. And [name of the journal's] approach is not to have such external funding that you depend on, but that all the resources that are of course necessary to keep a journal running come from the institutions involved. […] But the two technical editors are financed by my institute. Of course, I could also do completely different things with the positions and do real, I would say real, research. But that's the way it is. Well, that's academic freedom, that you can use the funds for the things that you think are right." (I-02, items 232-251)

By comparing the situation of this journal with the other journals of the sample, it becomes clear that the financial model is exceptional: in this case, an individual professor spends his resources to pay the staff that is necessary to run a journal, or, in other words, the common good of free access to published research and the possibility to submit manuscripts is produced by a decision of an individual.

*Professionalization*

One immediate effect of a high monetarization of tasks together with temporary unlimited funding, and large journal teams is a high degree of the professionalization, which is reported by all journals. It is most obvious in the case of two journals that apply the same pattern of division of labor and describe the journals' operations as well-structured processes, with clearly defined tasks for all members of the team involved. When compared with other journals of the samples that are not located in the upper right quadrant, the main differences are the unambiguousness with which tasks are assigned to specialized members of the journal team and the extraordinary extent of services the journal provides. Examples are extensive checks for quality traits, plagiarism and image manipulation that are performed by the editorial offices before transferring it into the peer review process (I-17, pos. 395-396), the provision of a repository for archiving pre-prints of submissions to the journal (I-17, pos. 458-466), and the amount of care that is spent on manuscripts that are accepted for publication and that is described as follows:

> "So, they do the format changes, they do the layout, they do copy editing. And I think that's one thing that also distinguishes us from other journals, we actually spend time to clean up the manuscript where it's necessary. So, we do proper copy editing, we correct, we revise grammar if it's necessary. And we revise the language and image, we handedly check. If someone says compound three, then we look at compound three and make sure it is actually compound three, and we make sure the images look nice, and we produce a nice layout. And the shape that the manuscript has in the end, we often get feedback actually, that the authors appreciate that we spend some of our resources and time in that as well. Because we want to be as inclusive as possible. We don't want language to be a barrier, within reason." (I- 17, pos. 422-441)

One unique feature of the two journals is, that, even though the work is distributed amongst specialized experts, all members of the journal team aside from the editor-in-chief, managing editors and the board are physically located at one place:

> So, we have our office spaces in [Name of a town]. And there's various departments and one department is the editorial office, which is a team that handles incoming submissions. And then there is a different department, which is the production team, which handles the manuscripts at a later stage. So, they are responsible for the format and for the layout. And for the copy editing. So, it's two different sets of people. (I-17, pos. 548-554)

Moreover, the two journals also use their own online editorial management system that serves their particular needs and is also an in-house development.

*Development of the journal*

Besides the degree of professionalization of the day-to-day routines, there is also a long-term effect of non-terminated funding that can be described as space for the development of the journal and the adaption to new requirements and needs. In sharp contrast to the situation of journals in the lower left quadrant, all editors of journals take a mid- or long-term perspective and formulate goals they aim to achieve. One journal prepares the replacement of the Online Editorial Management System OJS by Edit Flow[21] (I-05, pos. 543-556), while another aims at automating the production processes and to replace the currently paid work of the production by voluntary and unpaid gifts of the editorial team (I-02, pos. 465-469). In other words, to secure the long-term operation of the journal, the aim is to move the journal from the upper right quadrant into the upper left one. However, it is again the two charity-funded journals that do not only have enough resources to be spent on the journals' mid- and long term innovative but also implemented mechanisms for a constant collection of the feedback of the scientific community:

> "Again, this is all community driven, and we're constantly in contact with the community. If there were a need, we would consider this and we're constantly getting feedback. We have different events and workshops and are just in constant contact with the community and when there's a need, we would consider this." (I-17, pos. 184-188)

In the following quote it becomes clear that the journal team does not understand their role only as an implementor of ideas suggested by the scientific community but also as innovators that come up with own proposals and ideas and ask for feedback of the scientists:

> "So, every time we ask this question: 'Are you ready? Can we do open peer review?' But the chemistry community is-, they're just not ready, and they don't want it. And since we are a scholarly-led journal, we go with the wishes of the community, right." (I-17, pos. 508-513)

To summarize, the interviews with editors of Diamond OA journals from the upper right quadrant suggest that the position seems to be desirable: they have enough resources to cope with the day-to-day work, the division of work is defined by a high level of professionalization. Moreover, sufficient resources are available for the development and adaption of the journal towards the requests and needs of the scientific community to make it a really scholarly-led journal. Unfortunately, such amount of permanent funds is not available in all scientific disciplines and fields.

## 6.3 Lower-left quadrant: Individual leeway and struggle for resources

The journals of the lower left quadrant are characterized by a low degree of monetarization of tasks and small journal teams and are called 'small voluntary-run journals' by Bosman et al. (2021, p.104). The completion of tasks is in the first place gift-based and, because of the limited team size, individual contributions to the journal tend to be large. For the journals of the quadrant we will discuss four effects of the position: the limitations of gifts for the achievement of targets, the shortage of resources, the transfer of work within the journal teams and the long-term stability.

---

[21] https://editflow.org/ (accessed April 11 2023).

*Limitations of gifts for achievements of journals targets*

First, the gift-based character of contributions to the journal has consequences for shaping the processes within the journal. In such non-reciprocal gift systems, neither can a gift be demanded by the receiver, nor is the receiver in the position to determine *what* is given – both is up to the donor. For journals of the lower left quadrant, contributions are not standardized like in the case of the journal J-06 (the 'Miracle of the crowd' model). Therefore, frictions may appear between a donation and the requirements of the journal as perceived by the editors. One example is given by interviewee I-12, who would like to publish articles as soon as the peer review process is completed but is unable to do so because of the kind of technical support the journal receives.

> Q: "Do you publish the articles as soon as one is ready? Or is it by volumes and numbers?"
>
> I-12: "[…] No, we would like to. But that doesn't work technically. So unfortunately, we have to wait until the last article is here. I couldn`t stop there, well, I can`t tell this technical support from M. [name of a university], I WANT it to be done. But I`m grateful that they do it. And unfortunately, we have to accept that. Because they are not able to do that in any other way right now." (I-12, pos. 551-560)

In the quotation it becomes clear that the gift-based logic of contributions restricts the leeway of the editor: he may ask the donor for a different type of contribution – in this case changes of the configuration of the online editorial management system – but is unable to enforce it. This can only be imagined when work is paid but not in the context of a gift-based completion of tasks. A second example of limitations that arise from gift-based contributions refers to the speed in which the LaTeX formatting of articles is performed by an unpaid volunteer that does not meet the expectation of the editor:

> "And that`s what one of us is doing now, I haven`t mentioned him yet, that there is LT [name of the colleague]. He`s very good at it. We do that with LaTeX. He has worked his way into it, prepared semi-automatically by FE [name of the founding editor]. And he's doing it now, right? It has the big disadvantage that he also has other things to do and that sometimes it just takes a little longer than you would like." (I-07, items 350-362)

*Shortage of resources*

Resources that are necessary to run a Diamond OA journal are a prominent topic in the interviews with editors of journals that are located in the lower left quadrant, and time and money are often described as short. Therefore, a recurrent topic are tasks that cannot be tackled because of a lack of resources. A typical response to such kind of situations is that editors give priority to the short-term day-to-day operation of the journal and concentrate the available resources. This happens at the expenses of more long-term tasks.

> "I would have to sit down with IT, I would have to sit down with the graphics department, excuse me, but it's so simple in the end, I would have to sit down with one or two people from [Name of the Institute that owns the journal] […] We actually need time to do this restructuring - actually it can really be done within three or four days, I think, technically. Although technically, how do I know? Because I've learned that now in all this time, but-. And that's not how the university is organized. In the publishing house, I think it could be done in a kind of editorial conference, no idea, but. We don't. […] as I said, is as complicated as it is complicated and where I myself have hardly any room for maneuver." (I-11, pos. 983-1007)

The interviewee I-11 addresses the problems he encounters when trying to restructure his journal. Besides the lack of resources, a second problem is described: the journal and its editorial office are hosted at a university, an organization that has multiple missions (Schimank, 2001; Laredo, 2007; Engwall, 2020) and is not, like publishing houses, primarily dedicated to the publication of research. In such an organization, the restructuring of a journal involves a number of different departments and bodies. As a result, there is no blueprint on how to manage such a process. Therefore, not only the new structure of the journal has to be invented but also the process in which such a new structure can emerge.

In another interview it is again the development of the journal that can be thwarted because of lack of resources that appear to be a limiting factor:

> "We also keep discussing what we can improve. And then always try to optimize small areas. And so, we definitely have the room for maneuver. So that wouldn`t be a thing. It`s going to be difficult if we actually do it again, that is, if we need developer hours or something. So as long as we can do it and it`s our working hours, we have all the freedom. As soon as we would need someone externally, things would get tighter. So, you would have to think about it, should we apply for a third-party funded project?" (I-12, pos. 990-1000)

*Transfer of work and overburdening highly committed members*

The shortage of resources addressed in the previous section also affects the way in which workload is distributed in the journal team. A first mechanism that can often be found in the interviews is the *transfer of workload*. At its best, it happens reciprocally and helps all members of the journal team to balance their work for the journal with other obligations.

> "But we also do it a bit ad hoc. So, if someone says, 'it's really bad for me, I can't get anything done,' then someone else takes over. […]. And if I have a little more time, then I'll do another review from someone. He says, for example: 'Yes, I can't get around to it at all.' So that's a bit flexible. And there are definitely months when I do less. And some where I do more." (I-12, pos. 953-961)

However, the transfer of workload can also be unidirectional, typically from less involved team members to more strongly committed ones.

> "Okay, then we have to think about who could handle the editorial process. To be honest, in such special cases it often gets left up to the editorial managers. […] That's really work and also nerve-wracking. Because we don't want to overstrain our co-editors either. They all do it voluntarily, as do the reviewers. Nobody gets money for anything here." (I-11, pos. 403-409)

Here, the voluntary character of gift-shaped character of contributions becomes apparent. Given that there is the risk that the members of the board get overstressed (and may leave the journal as a consequence), the managing editor is reluctant to put pressure on them. As a result, a transfer of the workload takes place that also bears a risk: if it takes place on a regular basis, the capacities of the *highly committed volunteers may be overburdened.* An example can be found in the interview with the editor of J-09, a journal that changed its model from Diamond OA to a publication-based business model in search for resources. At the time of writing, the journal struggles with a situation that is characterized by good willing volunteers, lack of resources, and an overstrained editor-in-chief:

> "But at some point, what happened in every start-up, very late, stirred in me. That you have this feeling: 'You are left alone. You have to do it yourself, you can't rely on the DFG[22], you also can't rely on the APCs and not at all, you're just on your own. You had to make sure that you could manage it somehow. Whatever that means.' But the interesting thing is that you either stop there. Some stop. Or you keep going. Although without the support of these people who give their free time, I wouldn't even have been able to deliver." (I-09, pos. 1568-1577)

*Indispensable team members and long-term stability*

---

[22] Interviewee I-09 is referring to the Deutsche Forschungsgemeinschaft, a third-party funder from whom the journal previously received funding.

An effect of the gift-shaped contributions and small journal teams is that a journal may strongly rely on contributions of a single member of the journal team. Such dependencies are reported by all interviewees of the journals of the lower left quadrant, and two types can be distinguished. First, a team member can be indispensable because of a specific type of gift that no other team member can provide. Often such contributions are based on technical competencies, for example, the administration of a journal platform. Second, it is possible that the workload is focused on one dedicated editor who provides extraordinary-sized gifts that other team members are unable or not willing to contribute.

> "But I`ve been there since it was founded, and I`ve tried several times to be replaced. It`s difficult because at the end of the day I`m the only one who knows exactly how things are going, and that annoys me too. […] So, if I retire next year, for example, I can leave with a bang and then nobody will know how this IT system works. Or I invest even more time, although my whereabouts here is uncertain, to train someone. And I honestly think to myself, shit man, why would I do that? Yes, why should I do that? Yes? If I haven`t gotten anything for it so far, other than, I don`t know, I`ve helped 800 articles into the world- yes. So, don`t get me wrong, I`m not frustrated, I like doing it and I enjoy it too. But the question of what to do is definitely a question of resources." (I-11, pos. 983-1046)

Here, the first type of dependency also becomes apparent, as the editor has individual competencies regarding the IT system that the other members of the team lack. Aside from that, he was very much engaged in the reviewing and publication process of a large number of articles and assumes that he can only be replaced if a payment is offered to his successor. Both types of indispensability are strains for the long-term existence of a journal. If a highly engaged member or one with special competencies will leave the journal team, it may put the existence of the Diamond OA journal at risk.

## 6.4 Lower right quadrant

Finally, we turn to journals of the lower right quadrant that are characterized by a combination of a high monetization of tasks and small sized journal teams. The sample of these journals illustrate that the sources of funds that allow high monetization are diverse and include joint funding from the federal government and the federal states or from a learned society. However, a focus can be identified as five of the journals receive funds from the research institutions in which the editorial office of the journal is based. Another characteristic of the journals is that all of them collaborate with an infrastructural service provider that maintains an online editorial management system or at least the publication platform. In five cases, the service provider is a university library, while in two cases it is a commercial publisher. Regardless of the type of service provider, the tasks he is responsible for are all monetized. Therefore, the characterization of the journal as having a strong degree of monetization of tasks is at least in part a result of the collaboration with a technical service provider.

*Shortage of funds and low division of labor*

Although the monetization of tasks is relatively high, some of them, in particular those that are located close to the lower left quadrant, show indications of underfunding. This is the case if a task that should be part of the process of article production from the perspective of the editors cannot be performed because of a lack of monetary resources. Typically, such tasks are, on the one hand, not regarded as 'scientific' and, on the other hand, not part of the technical service providers' obligation. In the interview with I-20, the editing of articles is an example:

> "We can't afford good editing of our articles. And with English-language journals, it would make sense to have a native speaker for editing. Or at least to be able to have editing work carried out on a fee basis. Yes. That's hardly possible. We do that every now and then. But not all articles are editable. It would also make much more sense to have certain things processed by a research assistant. So, to set up the editorial staff better. But that is then a matter / also either a matter of the publisher or a matter of the faculty, or whatever. We work under financial conditions that make […] professional work hardly an option. We have to professionalize ourselves. In the different areas. Yes, we are amateurs in that sense. And 'amateurs' now does not refer to our competence, but to our resources." (I-20, pos. 624-640)

The call for a more professionalized attribution of tasks stands in sharp contrast to the situation of journals that are run by a small team. An example is the journal J-13, which collaborates with a university library for the hosting of the journal. The managing editor with whom the interview was conducted describes his situation when the software for the journal hosting was updated as follows:

"There was a change recently, I think to the version - was it four of the software? So, everything is new, it took a lot of work to change it because the structure had changed in part and the layout had to be redone. And nobody was or is here who does it then, except for me." (I-13, pos. 581-588)

Underfunding of a journal with a small journal team also has consequences also for tasks that occur irregularly. In this case, tasks are often *not attributed to someone who is qualified but to someone who is available*, a mechanism of attribution that also indicates low professionalization.

Regarding the long-term perspective, the interviews show similarities to journals of the lower left quadrant that are also characterized by a high personalization of tasks and indispensable team members. In the case of the journals of the lower right quadrant, such attributes are also mentioned and viewed as problematic. One interviewee explains that the journal ran into problems because of a chronic illness of the former editor-in-chief. As a result, the 'open section' of the journal was closed and it now publishes special issues only to reduce workload (I-1, pos. 744-759). Another interviewee who acts as the managing editor expresses his concern about what will happen to the journal now that he is ill (I-13, pos. 283-292), and a third one considers the increase of the journal team to reduce the dependency of the journal from individuals (I-04, pos. 1009-1025).

*Time limited third-party funding and the necessity of transformation*

When comparing the history of funding of these journals, another pattern becomes apparent. Three of the journals that now enjoy funding by their home institution have formerly received timely limited third-party funding for the build-up of the journal. Such kind of project-based funding is also known from other countries (Morrison, 2016). Having the results of the neighboring lower left quadrant in mind where some of the journals are located that formerly received third-party funding, two alternative development pathways seem to be likely when third-party funding expires: the journals' editors succeed in convincing their home institution to dedicate funds to their journal (or have the means to fund the journal themselves) or the journal moves into the lower left quadrant, in which the monetization of tasks is reduced. Actually, in the interviews with the journals of the lower right quadrant, which sustained a relatively large degree of monetization of tasks, the duration of funding is described as an unproblematic period, while the expiry of funding had a number of consequences. One of the interviewees whose funding for his journal had just ended at the time of the interview describes the situation as follows:

"But the situation in September is, of course, dramatic because the entire staff is falling apart, so to speak. That means, yes, we have to redistribute the work and N. [Name of a colleague] has a lot to do, I have a lot to do and, of course, that's really difficult. […] Because to make it very clear, to employ staff, you would need a new financier. […] But with this funding line, it was clear: three years and then it's over." (I-16, pos. 772-788)

The way in which the editors of the journal respond to the expiration of project funds is typical and includes the search for new sources for the acquisition of funds and, after the failure of this strategy, the restructuring of the division of labor.

"So we considered many options. Some of them - well, the idea actually came up, shouldn't we ask for publication fees? But even the administration of it would be in such a blatant disproportion to what would come out of it. And that's how we see it, we can't take money from the third-party funder for the development of the process and then convert it just because the funding is gone. […] But that means, of course, that is, basically maintaining the journal alone, the administration, that is a lot of work. We will then have to invest more time on the weekends by proofreading, which the 65% position has also done so far. And right, but now what will definitely be dead by September as of today, is any marketing activity. So there won't be any more resources for us to go out and say 'Hello Argentina, do you want to write something in English or book reviews here' because then we'll basically be dealing with managing the magazine. And, of course, we are concerned about what that means for further visibility." (I-16, pos. 792-812)

As explained in the quotation, the reactions are twofold: first, the workload to run the journal is redistributed among the highly committed members of the team and transferred from monetized to a gift-based completion of tasks. As already discussed for the journals of the lower left quadrant, the principle of such distribution of tasks may bear the risk to overstrain the capacities of the highly committed team members. Second, when resources are scarce, the editors tend to focus the journals' resources on tasks that are essential for getting the day-to-day work done. This is then at the cost of tasks that are more long-term oriented or with long-term effects like the decrease of the journal's visibility.

*Cooperation with technical partners*

A third characteristic of the journals of the quadrant is the cooperation with technical partners. Most of the interviewees state that they are satisfied with the collaboration and the division of labor between the editorial office and the service provider. An example is journal J-04, for which the editor-in-chief appreciates the collaboration:

"As far as the UB [Name of the university library] is concerned, they are certainly even more involved through the hosting of the OJS because we always have questions that go beyond our access options via the backend. We are very satisfied because we receive very good and very fast technical support from the university library. So, both in terms of technical issues related to the hosting of the OJS but also other issues, such as indexing the journal at Sherpa Romeo or at DOAJ. These are all things, we also get library expertise from the UB and they are happy to pass it on." (I-04, pos. 1122-1131)

However, the provider of infrastructural services may not always follow the inner logics and the priorities of science but may have their own orientation. As Schimank and Volkmann showed, publishers are guided by a scientifically finalized economic orientation and both orientations can be either strong or weak (Schimank & Volkmann 2012; Volkmann et al. 2014). In our sample, two types of service providers are represented and different types of frictions are reported. One of the journals collaborates with a commercial publisher in-line with the subscribe-to-open model. In this model, the income of the publisher depends on the willingness of enough libraries to keep up their subscriptions and to voluntarily support their Diamond OA model. In the case of the journal of our sample, this orientation of the publisher clashes with a careful and outstanding organization of the peer review process of the journal.

"But we are increasingly being put on the curb. And that's less the editors, but rather the publishers. Increasingly put on the curb because the publisher says yes, that these new models, subscribe to open and so on, are also associated with a certain risk for them. And we simply can no longer afford to publish very specific issues with a delay of several months. And that`s why they insist very intensively that we deliver our contributions on time. And, of course, that has very specific consequences. Right down to the editorial staff and publishers. Yes. A lot of things have to be simplified, accelerated and so on and so forth. […] So they have no control in that sense. They don`t even claim that. But they are based on the output. It's a kind of output control. Yes. But no process control or anything. No." (I-20, pos. 871-886)

A second type of clash can be observed between a scientific orientation of the editorial board and the orientation of a library that acts as an infrastructural service provider for the journal. In the case of the journal J-16, tensions occurred between the goals of the editors to build-up a publication channel for an interdisciplinary field in the social sciences and the orientation of the library to develop a hosting service for journals that could be rolled out for other journals with editorial offices at this university.

> "So, we were always treated by the [Name of the university library] as an exemplar for subsequent journals, so to speak. So, they wanted to try it out on us, how it works. We were the first journal that really got off the ground and then we were supposed to always serve like that, which also had clear disadvantages. We are now setting the standards for everyone else, so to speak. And then we weren`t just seen as J-16, but this thinking was always in the background, yes, that's tailor-made for you, but above all, you have to be a blueprint now. What we do with you must potentially be transferrable to everyone else. That has sometimes caused us difficulties." (I-16, pos. 689-704)

In other words, the conflict between the editors and the library is that between standardization of a service vs. tailor-made support. Given that small independent journals are diverse (Morrison, 2016), the editors look for individual solutions that fit to the particularities of their journal, while the library aims to develop a solution that is suitable for all Diamond OA journals at their institution that may look for a technical partner in future.

## 7. Conclusion

This paper aimed to map the German Diamond OA journal landscape in two complementary ways. The bibliometric analysis provides a top-down picture about the characteristics of the whole landscape, while maps based on the qualitative results of expert interviews focus on resource-related differences that effect the situations of the journals. Returning to the initial question about the capability of the Diamond OA model, we will summarize the most important results as follows: with the existence of a substantial number of Diamond OA journals in the social sciences and humanities, the German OA journal landscape supports existing evidence (Bosmann et al., 2021, Hahn et al., 2022) that the publishing models work in these fields. The focus in the two fields and a tentative update in the natural sciences, engineering and technology, agricultural sciences, and medicine does not imply that the publishing model does not work but that other OA types are more appealing or better established in these fields (Severin et al., 2020). Moreover, the analysis of the size of Diamond OA journals suggests the model works for small to mid-sized journals. This result should not be interpreted that Diamond OA journals are marginal or of subordinate importance. On the contrary, they are often important (and sometimes the only) means to publish research in particular in small specialties, as a number of the journal titles suggest. However, the lack of large journals with an article output > 1,000 shows that – at least the German Diamond OA landscape – is far from challenging the established publishing industry. The fact that the average publication output of Diamond OA journals that are published by commercial publishing houses is larger than those published by institutions indicates that the larger the publication output is the larger is the need for collaboration with publishing professionals. Evidence from the expert interviews shows that there are functional equivalents to commercial publishers like institutions or in-house publisher-alike structures.[23] Finally, the bibliometric study showed that a considerable number of 23 Diamond OA journals (or roughly 5% of the total number) have not published any publication during the last two years and are therefore considered as being discontinued. Such a high number indicates the German Diamond OA journals develop dynamically.

The in-depth qualitative analysis based on a survey and interviews with experts proved that the field is differentiated and the two dimensions 'monetization of tasks' and 'size of the journal team' help to understand important differences in the field. The analysis of the interviews with journals in the two upper quadrants conveyed that there are sustainable models on which also journals with a mid-size publication model are based. The different degrees of monetization points to two alternatives: the community-driven journal (or the 'miracle of the crowds'-model) and well-funded professional journals, in which not only infrastructural or service-oriented tasks are paid for but also the more scientifically tasks in editorial offices, like the organization of the peer review process, are supported with monetary resources.

---

[23] Most prominently in the case of J-17.

Journals in the two lower quadrants show some characteristics that question their long-term stability: in the lower left quadrant, the shortage of funds tends to result in a low division of labor and we find instances of a transfer of work to highly committed members. Given that they are often indispensable because of the workload they shoulder, such transfers are problematic as they bear the risk to overburden them. Although gift-like contributions seem to be at first glance a likable mode of running a journal, it also has a problematic side from the perspective of the editors. A possible donor does not only decide whether or not to contribute but also on how to contribute. This may result in frictions between the gift given by donors and the requirement of the journal as perceived by the journal editor. Finally, journals in the lower right quadrant are often financed by third-party funds that result in a necessity for a transformation. In contrast to a large variety of resourcing Diamond OA journals (Bachmann et al., 2022), the interviews show that there are two main pathways on how to proceed with a Diamond OA journal after third-party-funding has expired: to convince the home institution in which the editorial office is located to fund the journal or to reduce the monetization and increase the degree of gift-based completion of tasks. The second option moves a journal into the direction of the lower left quadrant.

Finally, we would like to conclude with a personal impression. The interviews have demonstrated that the editors are highly committed to their journal and, often in contrast to their perception of the attitude of the community they serve, convinced of the value of a publishing model that comes without financial barriers for both authors and readers. However, if we want Diamond OA journals to succeed, it is necessary to develop funding mechanisms that are not limited in time. Such financial mechanisms already exist for infrastructures but are underdeveloped for the editorial work. The capabilities of the editors together with contingencies at home institutions should not be decisive for the stability of a Diamond OA journal that proved to be successful in scientific terms.

## Literatur